# WARM GAS AT HIGH REDSHIFT

*Clues to Gravitational Structure Formation from Optical Spectroscopy of Lyman Alpha Absorption Systems*

MICHAEL RAUCH

*Astronomy Department*
*105-24, California Institute of Technology*
*Pasadena, CA 91125, USA*

**Abstract.**
  We discuss the effects of gravitational collapse on the shape of absorption line profiles for low column density (N(HI) < $10^{14}$ cm$^{-2}$) Lyman $\alpha$ forest clouds and argue by comparison with cosmological simulations that Lyman alpha forest observations show the signs of ongoing gravitational structure formation at high redshift. The departures of observed line profiles from thermal Voigt profiles (caused by bulk motion of infalling gas and compressional heating) are evident from the results of profile fitting as a correlation in velocity space among pairs of components with discrepant Doppler parameters. This correlation also allows us to qualitatively understand the meaning of the Doppler parameter - column density (b-$N_{HI}$) diagram for intergalactic gas.

  A part of this conference was devoted to the prospects of detecting neutral hydrogen at high redshift as tracers of the gaseous large scale structure (contributions by Braun, de Bruyn, Ingram, Swarup, and Weinberg). A successful detection with radio-astronomical techniques depends on the gas being in a state of high HI column density (N(HI) > $10^{18}$cm$^{-2}$) and low temperature. We have reason to believe that gas condensations where such conditions prevail are rather rare in terms of geometric cross section, volume filling factor, and probably also in terms of the total fraction of baryonic matter they represent. Here we take a complementary point of view and ask what we can learn by observing more typical low column density, warm (T$\sim$ a few $\times 10^4$ K) gas at high redshift. To study such tenuous gas condensations (the so-called Lyman $\alpha$ forest) we need to look for the absorption imprinted by intervening gas onto the spectrum of a strong background source, an AGN or QSO. Typically, with a large optical telescope like the 10m Keck neutral hydrogen column densities down to $10^{12} cm^{-2}$ can be measured in a few hours of observing time over a simultaneous redshift range of $\Delta z \sim 1$, in front of a 17-18th magnitude QSO.

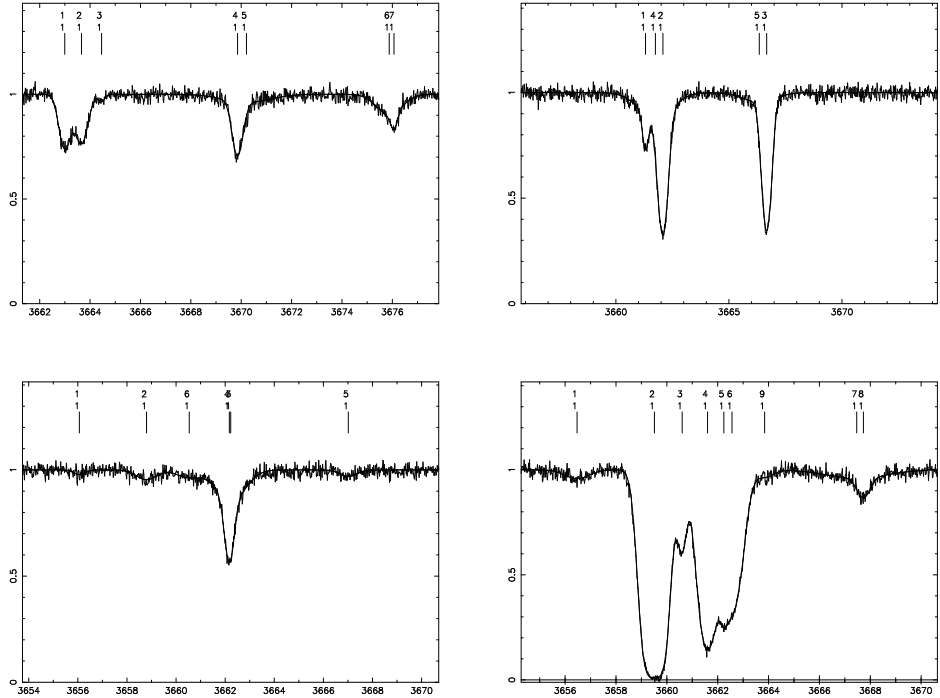

*Figure 1.* Spectral regions from lines-of-sight through the z=2 simulation of Miralda-Escudé et al., showing absorption lines departing from single Voigt profiles. Several of the clouds show asymmetries and are better fitted by very close pairs of a broad and a narrow component.

What causes the Lyman $\alpha$ forest phenomenon ? Recent numerical simulations of gravitational structure formation within the cold dark matter scenario (including the effects of gasdynamics and ionizing background radiation) have brought us closer to an answer to this question (Cen et al. 1994; Weinberg, this volume; Zhang et al, 1995; Miralda-Escudé et al. 1995). According to these experiments gravitational collapse of baryonic matter produces extended (length scale of order 1 Mpc) condensations of gas giving rise to absorption phenomena very similar to the observed Lyman alpha forest. Typical low column density Lyman $\alpha$ clouds appear to be sheet-like or filamentary structures with relative overdensities of $\sim 1 - 10$). During the epoch accessible to observation collapse of gas and accretion continue – we are watching gravitational structure formation in situ.

If this picture is right, what is the observational signature of this formation process ? Assuming the clouds are really objects in a phase of collapse then bulk motion and compressional heating of the infalling gas should affect the absorption line profiles of these clouds, producing departures from a Maxwellian velocity distribution. Line shapes are then no longer well represented by a Voigt profile (which would characterize a static, homogeneous temperature phase). Nevertheless, fits with multiple, pure Voigt profile models to entire high resolution Lyman $\alpha$ forest spectra (e.g. Carswell et al. 1991) have yielded excellent results statistically indistinguishable from the data, a statement which remains true even at the very high signal-to-noise ratios (up to 100 or more) achievable with the Keck telescope (e.g. Tytler et al. 1995). Thus, if a decomposition in terms of Voigt profiles works but the real absorption line shapes individually depart from such profiles, all the information about the physics must be contained in the correlations among various parameters (redshift, Doppler parameter, HI column density), analogous to a Fourier decomposition of a periodic function (though the analogy is limited, as Voigt profiles are not orthogonal base functions).

Figure 1 shows four regions from artificial spectra created from the simulation of Miralda-Escudé et al. (1995) for redshift 2. Close inspection reveals a number of lines of intermediate strength (column densities below $10^{14}$cm$^{-2}$ consisting of a narrow central core surrounded by broad, often asymmetric wings. Comparison with the physical parameters of the simulation shows that the wings are caused partly by the bulk motion of infalling gas and partly by temperature gradients due to compressional heating, although in individual cases it is difficult to disentangle the contributions to the width from temperature and bulk flow.

Modelling the simulated lines by Voigt profiles at an assumed signal-to-noise ratio of 50 often requires two (occasionally more) components, close together in velocity space, a narrow one for the core and a broad one representing the non-maxwellian wings. Thus, one spectral signature of gravitational structure formation would be an anticorrelation of Doppler parameters for profile pairs at very small (0-30 km/s) separation. In figure 2 we plot the fraction of line profile pairs with Doppler parameters discrepant by more than a certain factor R, i.e. $|log(b_1/b_2)| > log(R)$, as a function of the velocity splitting between the components of the pair. Data in the upper diagram are from the Miralda et al. z=2 simulation. As expected, there is a strong excess of pairs with very discrepant Doppler parameters at the smallest splittings. Dividing the sample into subsets with different column densities it can be shown that this signal is dominated by lines with column density N(HI) <$10^{13.5}$cm$^{-2}$. If the combination of narrow and broad components were an artifact in that we would be substituting a broad component for a number of unresolved normally narrow ones in a blend, then we would expect to find the higher column density, more clustered absorption systems to provide the bulk of the correlation signal, contrary to what we are seeing here. The lower half of fig.2 shows the same statistic for the actual observations from the work of Hu et al. (1995). The mean redshift here is higher ($<z> \sim 2.86$), and the line fitting was done somewhat differently (the deficit in the zero velocity bin is likely to be an artifact), but the overall picture, the discrepancy among Doppler parameters, increasing with decreasing velocity separation, as a signature of non-Voigt profiles, is similar. Obviously, the observed absorption line shapes are consistent with the expected line profiles from gravitational structure formation scenarios.

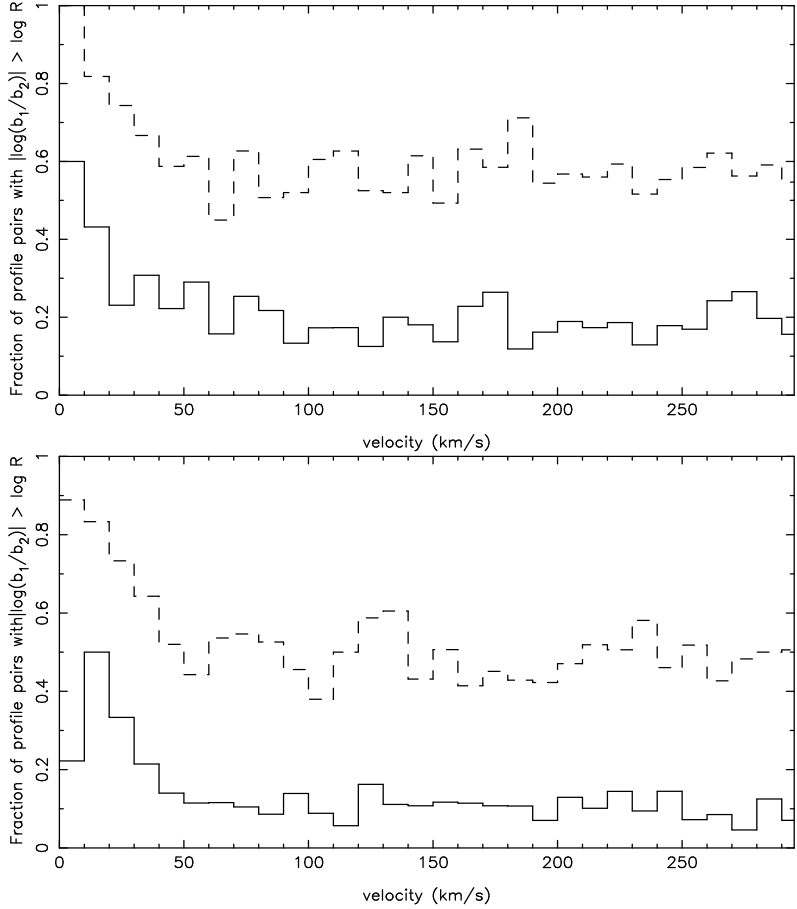

*Figure 2.* Fraction of absorption line pairs with Doppler parameters different by more than a factor 3 (solid line) or a factor 1.5 (dashed line), versus the velocity splitting between the components, for absorption lines at redshift 2 from the Miralda-Escudé et al. model. Apparently, at small splittings an absorption line can be decomposed into two close components with very discrepant Doppler parameters, indicative of a narrow line core with broad wings. Below, the same statistic for the real data from the Hu et al. (1995) Keck sample. There is indeed a signal similar to the one from the simulation.

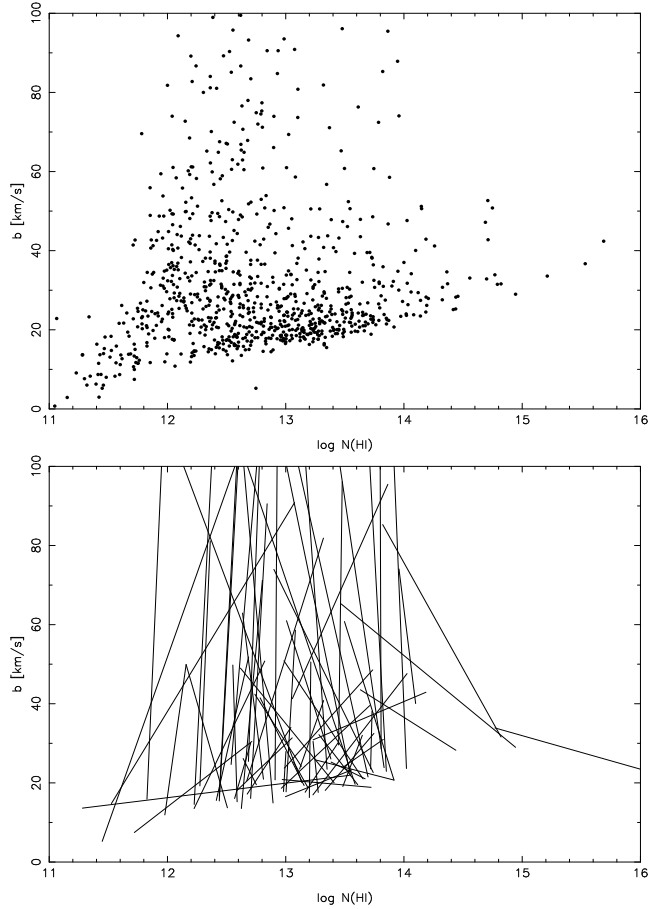

Figure 3. Above: Doppler parameter b versus HI column density N(HI) for Ly$\alpha$ from 100 lines of sight through the Miralda et al. 10 Mpc box at z=2. For clarity only the lower 100 km s$^{-1}$ are shown. Main features also seen in real data are: (1) the absence of lines below a minimum Doppler parameter for a given N(HI), a weak positive correlation between b and N(HI), and a significant population of large Doppler parameter components at lower column densities (b's ranging up to several hundred km s$^{-1}$). Below: b-N diagram of pairs of lines with velocity splitting < 20 km s$^{-1}$ corresponding to the first two bins in figure 2. Each pair (i,j) is represent by a straight line, with [(b$_i$, N$_i$),(b$_j$, N$_j$)] as its endpoints. At a velocity separation this close, the nearest neighbour of a narrow line close to the photoionization cutoff (lower envelope) is usually a broad companion (often with Doppler parameter larger than 100 km s$^{-1}$) rather than a component of random Doppler parameter.

Of course, this is not a proof that the structure formation scenario for intergalactic gas clouds is correct. We could postulate the (ad hoc) existence of unresolved clustering of weak lines on very small scales to explain the presence of broad wings, but to emulate the large width of the broad components would require a rather contrived column density distribution for the cluster components - currently there is no physical justification for such a picture. The existence of galactic winds could be an alternative physical model which cannot be discussed on the basis of line profiles alone as we cannot distinguish between outflow or infall, but the presence of absorption coherent over Mpc scales (e.g. Dinshaw et al. 1995) makes this possibility less likely. In any case our above result is probably the first direct observational evidence we have of these objects being in a dynamic state.

The above result also helps to clarify the longstanding issue of the meaning of the supposedly "supra-thermal" Doppler parameters in the b - N(HI) diagram (figure 3). From what was said above it is clear that many of the broad low column density lines in this distribution are actually not independent systems but the bulk motion- and temperature-broadened wings of absorption lines from gas clouds with more quiescent gas in the center, the temperatures of which are consistent with photoionization heating plus some contribution from the potential well in which the gas has collapsed. Together with a second population of low column density clouds in voids with line profiles dominated by bulk motion (Miralda-Escudé et al. 1995) these broad wings account for the high Doppler parameters in the b-N diagram. It is hoped that the future analysis of large Lyman $\alpha$ forest data samples at different redshifts, in connection with gasdynamical simulations will teach us more about the thermal history of cosmologically distributed gas.

I thank my collaborators on various projects, Bob Carswell, Renyue Cen, Len Cowie, Esther Hu,Tae-Sun Kim, Jordi Miralda-Escudé, Jerry Ostriker, Wal Sargent, Tony Songaila for permission to quote from some of the results prior to publication.